# Numerical study of energy diffusion in King models

Tom Theuns
*Department of Physics, Nuclear Physics Laboratory, Keble Road, Oxford OX1 3RH, UK*
*tom.theuns@physics.oxford.ac.uk*

**ABSTRACT**
The energy diffusion coefficients $D_n(E) \equiv \langle \Delta E(E)^n / \Delta t \rangle$ ($n = 1, 2$) for a system of equal mass particles moving self-consistently in an $N$-body realisation of a King model are computed from the probability per unit time, $\mathcal{P}(E, \Delta E) dE d\Delta E$, that a star with initial energy $E$ will undergo an energy change $\Delta E$. In turn, $\mathcal{P}$ is computed from the number of times during the simulation that a particle in a 'state' of given energy undergoes a transition to another state. These particle 'states' are defined directly from the time evolution of $E$ by identifying them with the event occuring between two local maxima in the $E(t)$ curve. If one assumes next that energy changes are uncorrelated between different states, one can use diffusion theory to compute $D_n(E)$. The simulations employ $N = 512, 2048, \cdots, 32768$ particles and are performed using an implementation of Aarseth's direct integrator $N$-body1 on a massively parallel computer. The more than seven million transitions measured in the largest $N$ simulation provide excellent statistics. The numerically determined $D(E)$'s are compared against their theoretical counterparts which are computed from phase-space averaged rates of energy change due to independent binary encounters. The overall agreement between them is impressive over most of the energy range, notwithstanding the very different type of approximations involved, giving considerable support to the valid usage of these theoretical expressions to simulate dynamical evolution in Fokker-Planck type calculations. Even so, diffusion, as judged from these measurements of the diffusion constants, is stronger than expected from theory, both in core and outer halo, by a factor up to two, rather independent of particle number. The experimental $D$'s obey very well the expected scaling $\propto N/\ln \Lambda$ with particle number $N$.

## 1 INTRODUCTION

The energy per unit mass $E = v^2/2 - \phi$ of a star moving with velocity $v$ through a stellar system is not conserved when the potential $\phi$ changes in time. Temporal changes in $\phi$ are due to the whole spectrum of modes in the system which range from collective modes (which also occur in the limit $N \to \infty$, e.g. Weinberg 1993), over multiple encounters all the way down to binary encounters. This redistribution of energy over the different particles may lead to structural changes of the system.

Following Spitzer (1987), the dominant contribution to the changes in particle energy for a system in dynamical equilibrium comes from the cumulative effects of many distant binary encounters, each changing $E$ by a small amount and causing $E(t)$ to perform a random walk in energy space (see Spitzer 1987, p. 29 for a discussion). In this case it is appropriate to describe the rate of change of $N(E,t)dE$, the number of particles with energy between $E$ and $E + dE$, by a diffusion equation.

A theoretical expression for the diffusion coefficients in a system of particles with $1/r^2$ forces and characterised by an arbitrary distribution function is given by Rosenbluth et al. (1957). The derivation assumes two-body encounters are independent and sums the contribution from encounters of given impact parameter $b$ and impact velocity $v$. Unfortunately, the resultant expressions diverge logarithmically in the gravitational case due to the contribution of distant encounters because, unlike in the plasma case, such distant encounters are not screened. These diffusion coefficients– with an imposed cut-off, the Coulomb logarithm– are widely used in Fokker-Planck type calculations (see e.g. Chernoff and Weinberg (1990) or Spitzer (1987) for references).

In a real system, a particle is undergoing several encounters with a range of impact parameters *simultaneously*. Consequently, the approximation where one just adds the contribution from all impact parameters for pure two-body motion is questionable. Indeed, for the dominant encounters with intermediate impact parameter, the description of the motion as being a pure two-body encounter is likely to be an oversimplification. In addition, the description likely overestimates the importance of more distant encounters, both in homogeneous and in concentrated systems. In a homogeneous system for example, it would seem that the many encounters with impact parameter $b \ll l$ (for some suitably chosen $l$) perturb the orbit of a particle under consideration to such an extent that encounters with $b \gg l$ are effectively terminated, thereby quenching the contribution from encounters with such larger impact parameters. This argument is even stronger in a concentrated system: consider encounters with large $b$ between a core and a halo particle. The dynamical time scale of the core particle $t_{\rm dyn} \approx 1/\sqrt{\rho}$, with $\rho$ the density at the position of the core particle, may be much smaller than the typical encounter time $t_{\rm enc} \approx b/v$, with $v$ the relative velocity. Again, this leads to a quenching of the contribution of encounters with large impact parameter. In practice, ignorance of the appropriate value of the impact parameter of encounters that still contribute to the diffusion process is hidden in the value of the Coulomb logarithm used.

Similar considerations led Chandrasekhar (1941) to consider a different formulation of the problem based on the notion of particle 'states'. A particle is subject to the fluctuating gravitational field due to all the other particles causing $E(t)$ to perform a random walk. The force $F(t)$ acting on a particle is correlated over short times but uncorrelated over longer times, due to the chaotic nature of the system. A particle is considered to be in a particular state as long as $F$ is strongly correlated in time. It may then undergo a transition to another state and $F$ will not be correlated between these different states. The rate of diffusion in energy space of the particle can now be computed from knowledge of the average lifetimes of states $T(F)$ and the probability distribution $W(F)$ of a given force acting during a state. Maybe somewhat surprisingly, the resultant form of the expression for the diffusion coefficient is identical to that of the other formulation based on independent binary encounters, and their numerical values differ only by a factor $\approx 1.11$ (Chandrasekhar 1941).

The aim of this paper is to evaluate the diffusion coefficients $D_n$ ($n = 1, 2$) as a function of $E$ from a direct $N$-body simulation by studying the properties of the random walk in energy space for particles of given energy in a similar spirit as Chandrasekhar's theory of states. We define the states of the particle from properties of the $E(t)$ curve and gather statistics on the lifetime and transition probability of states to compute $\mathcal{P}(E, \Delta E)$, the probability per unit time that a particle of energy $E$ undergoes a transition to another state of energy $E + \Delta E$. Comparison can then be made between the numerically determined $D$'s and their theoretical counterparts as used in Fokker-Planck calculations. The advantage of the numerical method over the analytical derivation is that the full non-linear dynamics is treated consistently, properly taken into account the quenching of distant encounters, the effect of interactions during simultaneous encounters and the possible contribution from collective modes.

The diffusion coefficient $D_2(E)$ can be used to define the relaxation time $T_E(E)$ of the system (Eq. (20) below). Previous measurements of $T_E$ from $N$-body simulations have used a variety of other methods (see Huang et al. 1993 for more details), e.g. the rate of energy exchange between different mass components, the mean-squared energy change of equal mass particles and the measurement of deflections of test stars moving through $N$ field stars.

The rest of this paper is organised as follows: after defining the diffusion coefficients and giving their standard theoretical expressions in section 2, we describe the experimental set-up (3) and the experimental definition of the $D$'s (4). These are compared in section 5. The relative merit of the proposed method for measuring diffusion coefficients and the reason for the (small) discrepancies between theoretical and numerical measurements is discussed. Finally, the paper is summarised.

## 2  THEORETICAL DESCRIPTION

### 2.1  Theoretical energy diffusion coefficients

Spitzer (1987 and references therein) gives the standard expressions for the diffusion coefficients based on assuming independent binary encounters. Let $N(E,t)dE$ be the number of stars in a system with energy in the range $E$, $E + dE$ at time $t$. The encounter term in the Fokker-Planck equation is written as (Spitzer 1987, Eq. (2-71))

$$\left(\frac{\partial N(E,t)}{\partial t}\right)_{\rm enc} = -\frac{\partial}{\partial E}\left(N(E,t)D_1(E)\right)$$
$$+ \frac{1}{2}\frac{\partial^2}{\partial E^2}\left(N(E,t)D_2(E).\right)$$
$$+ \cdots. \qquad (1)$$

The diffusion coefficients $D$ can then be computed from phase-space weighting the energy changes $\langle \Delta E^n/\Delta t\rangle$ ($n = 1, 2$) undergone by a star per unit time. The phase-space weighted average $\langle A \rangle_V$ of a quantity $A$ is (Spitzer 1987, Eq. (2-80)): $\langle A \rangle_V \equiv \int_0^{r_{\max}} A\, vr^2 dr/p$, with $p$, the phase-space volume accessible per unit interval of $E$, given by Eq. (4) below. The expressions for $\langle \Delta E^n/\Delta t\rangle$ are given by Eqs. (2-50, 2-51) in Spitzer (1987). (We write $\langle \Delta E/\Delta t\rangle$ instead of Spitzer's $\langle \Delta E \rangle$ to stress these are energy changes per unit time.) If the background stars have masses $m_f$ then the diffusion coefficients for a star of mass $m$ are given by:

$$D_1(E) = 16\pi^2 m_f^2 \ln \Lambda \left(\int_E^\infty f(E_f)\, dE_f \right.$$
$$\left. - \frac{m}{m_f p(E)} \int_{-\infty}^E f(E_f)\, p(E_f)\, dE_f \right) \qquad (2)$$

and

$$D_2(E) = 32\pi^2 m_f^2 \ln \Lambda \left(\frac{1}{p(E)}\int_{-\infty}^E f(E_f)\, q(E_f)\, dE_f \right.$$
$$\left. + \frac{q(E)}{p(E)}\int_E^\infty f(E_f)\, dE_f\right). \qquad (3)$$

Here

$$p(E) \equiv \int_0^{r_{\max}(E)} (2(E-\Psi))^{1/2}\, r^2\, dr \qquad (4)$$

$$q(E) \equiv \frac{1}{3}\int_0^{r_{\max}(E)} (2(E-\Psi))^{3/2}\, r^2\, dr, \qquad (5)$$

and $r_{\max}(E)$ is the maximum radial distance a star of energy $E$ can wander and $\ln \Lambda$ is the 'Coulomb logarithm'. One has $\ln \Lambda \approx \ln p_{\max}/p_{\min}$ where $p_{\max}$ and $p_{\min}$ denote maximum and minimum impact parameter for binary encounters, respectively. The distribution function $f$ occurring in these expressions is normalised such that its integral over phase-space equals $N$. The collision term of the Fokker-Planck equation (2-86) in Spitzer (1987) is obtained upon substitution of Eqs. (2) and (3) into (1). The diffusion coefficients for a system of equal masses are given by:

$$
\begin{aligned}
D_1(E) &= \frac{16\pi^2 M^2 \ln \Lambda}{N} \left( \frac{1}{N} \int_E^\infty f(E_f) \, dE_f \right. \\
&\quad \left. - \frac{1}{N p(E)} \int_{-\infty}^E f(E_f) p(E_f) \, dE_f \right)
\end{aligned}
\qquad (6)
$$

and

$$
\begin{aligned}
D_2(E) &= \frac{32\pi^2 M^2 \ln \Lambda}{N} \left( \frac{1}{N p(E)} \int_{-\infty}^E f(E_f) \, q(E_f) \, dE_f \right. \\
&\quad \left. + \frac{q(E)}{N p(E)} \int_E^\infty f(E_f) \, dE_f \right),
\end{aligned}
\qquad (7)
$$

with $M = Nm$.

### 2.2 Coulomb logarithm

The Coulomb logarithm $\ln \Lambda$ occurs because of the divergence of the diffusion coefficients due to the cumulative effects of many distant encounters. In comparing the numerical $D$'s, defined in the section 4.3 below, with the theoretical expressions given previously, we will take $p_{\max} = r_c$ (Spitzer 1987, p. 28), with $r_c$ the core radius of the model, although other choices could be made as well (e.g. Spitzer (1987, p. 30) and Farouki and Salpeter (1994) argue for $p_{\max}$ equal to the half mass radius). In a system without smoothed gravitational forces one usually takes $p_0$, the impact parameter causing a 90° deflection (Eq. (16)), for the minimum impact parameter. This gives:

$$\ln \Lambda = \ln \frac{r_c \, \sigma(0)^2 \, N}{2M} \equiv \ln(\lambda_1 N), \qquad (8)$$

with $\sigma(0)^2$ characterising the central velocity dispersion.

In the numerical simulations presented here, gravitational forces are smoothed with $\epsilon = 0.5d$ (see Eq. (13) below) with $d$, the average central interparticle distance, given by Eq. (14) below. Substituting $\epsilon$ for the smallest impact parameter $p_{\min}$ gives the Coulomb logarithm appropriate for the numerical simulations:

$$\ln \Lambda = \ln \frac{r_c}{\epsilon} \equiv \ln(\lambda_2 N^{1/3}). \qquad (9)$$

Numerical values for $\lambda_1$ and $\lambda_2$ applicable to the King models studied in section 3.1 are:

$$
\begin{aligned}
\lambda_1 &= 0.028, \quad \lambda_2 = 0.019 \quad \text{for } W_0 = 9 \text{ and} & (10) \\
\lambda_1 &= 0.196, \quad \lambda_2 = 0.197 \quad \text{for } W_0 = 3. & (11)
\end{aligned}
$$

## 3 EXPERIMENTAL SET-UP
### 3.1 N-body model

All results presented here come from simulations of $N$-body realisations of King models (e.g. Binney and Tremaine 1987, henceforth BT87, p. 232). King models form a one parameter family of 'lowered isothermal' models whose distribution function $f$ has a sharp cut-off at the 'tidal' energy $E_0 < 0$. They can be characterised by the ratio $W_0 \equiv \Psi(0)/\sigma^2$ of the central potential $\Psi(0)$ over a parameter $\sigma$ characterising the velocity dispersion. Given the run of density, potential and velocity dispersion for a given $W_0$, a particular $N$-body realisation of this King model is made using a random number generator. For small particle numbers, the resultant system may be slightly out of equilibrium due to small $N$ statistics. In addition, the equations of motion are integrated by softening the gravitational force (section 3.3) which causes the initial state of all $N$-body realisations to be slightly out of equilibrium. The dynamical (or crossing) time used in the following is defined in the usual way as $t_d \equiv t_{cr} \equiv M^{5/2}/(2|E_T|)^{3/2}$, with $M$ and $E_T$ the total mass and total energy of the cluster, respectively. (In the 'standard' $N$-body units, where $M = G = -4E = 1$, $t_d = 2\sqrt{2}$.) Here and in the following we take the gravitational constant $G = 1$.

### 3.2 N-body code

The Newtonian equations of motion were integrated using Aarseth's $N$-body1 code (Aarseth 1985) in 15 digit precision using force smoothing and not including regularization. $N$-body1 is a high-order scheme which uses Newton divided differences to compute e.g. the force $F_i(t)$ on particle $i$ as a function of time $t$ as a Taylor expansion in $t$ including terms up to the fourth order. Such high-order expansions are also used to update the positions and velocities of particles using individual time steps $\delta t_i$, which are computed from (Aarseth 1985, Eq. (9)):

$$\delta t_i^2 = \eta \, \frac{|F||F^{(2)}| + |F^{(1)}|^2}{|F^{(1)}||F^{(3)}| + |F^{(2)}|^2}, \qquad (12)$$

where $F^{(i)} \equiv d^i F(t)/dt^i$ and $\eta = 0.03$ is an accuracy parameter.

This code was implemented on a massively parallel computer, the 8192 processor Connection Machine 2 at the Scuola Normale Superiore di Pisa. Parallelism was exploited in computing the force $F_i$ on particle $i$ due to all other particles $j$ in parallel using the FORTRAN 90 intrinsic function SUM. A catalog scheme was used to group particles with time steps equal to within a factor of two in bins. Such binning allows parallelism in updating the Newton divided differences and in addition significantly improves interprocessor communication efficiency. The resulting code runs at near 380 Mflops, about 25% of the theoretical peak performance, with typically $\approx 97\%$ of CPU time spent in pure force evaluation (see Theuns and Rathsack 1993 for more details).

### 3.3 Force calculation

The force $F_{ij}$ on particle $i$ due to particle $j$ is computed using direct summation and is softened according to:

$$F_{ij} = \frac{r_j - r_i}{((r_j - r_i)^2 + \epsilon^2)^{3/2}}, \qquad (13)$$

since otherwise large integration errors occur in the absence of regularization, due to the singularity at $r_i = r_j$. The size of the applied smoothing can be compared with several scales in the cluster: the average interparticle distance at the centre $d$, the semi-major axis $a$ of a binary just on the division line between being soft and being hard, (e.g. BT87 p. 534, $P_{\rm hard}$ being the period of this binary), and the impact parameter that causes a 90° deflection, $p_0$ (Spitzer 1987, Eq. (2-5)):

$$d = (m/\rho(0))^{1/3}, \qquad (14)$$
$$a = m/2v_m^2, \qquad (15)$$
$$p_0 = 2m/v_\infty^2, \qquad (16)$$

where $\rho(0)$ is the central density, $v_m$ the local velocity dispersion and $v_\infty$ the velocity at infinity for the two-body encounter leading to a 90° deflection. The dependence on $N$ can be brought out by defining a linear dimension $R$ of the model from $\rho(0) = M/(\frac{4\pi}{3}R^3)$ and by writing $v_m^2 \approx 2M/R$ and in addition substituting $v_\infty$ by $v_m$. This translates Eqs. (14–16) into $d = (4\pi/3)^{1/3}R/N^{1/3}$ and $a = p_0/4 = R/4N$. In the simulations presented, $\epsilon = d/2 \propto N^{2/3}p_0$, from which it is clear that strong encounters are completely suppressed for large N.

## 4 EXPERIMENTAL DESCRIPTION

### 4.1 Basic assumption

In the following we consider a given particle to remain in the same state in the time interval between two local maxima of the $E(t)$ curve of that particle. We identify the energy $E$ at the first of the two maxima as the state's 'energy' and the duration $\delta t$ of the interval as the state's lifetime. At the second maximum the particle makes a transition to a new state with in general different energy and lifetime. In the following we show how to compute the $D$'s from knowledge of the state's energies, lifetimes and transition energies.

This heuristic definition of a state implicitly assumes that energy fluctuations cease to be correlated between successive energy maxima, due to the chaotic nature of the $N$-body orbits. The typical lifetimes of the states defined in this way turn out to be of the order of the ratio of the interparticle distance over the local velocity dispersion, in line with Chandrasekhar's (1941) estimate.

The prime advantage of this definition is that the sampling rate used to determine the $D$'s is adapted automatically to the particle's rate of energy change: a particle undergoing frequent energy changes (a core particle) will be sampled at a higher rate than a particle undergoing fewer energy changes (a halo particle), thereby greatly improving the statistics on transitions between states. In addition, the expected systematic energy change, due to the diffusion process itself, is negligible with respect to the particle's energy over the lifetime of the state, so that the diffusion process can actually be studied as a function of the particle's energy. Finally, this method allows one to measure the diffusion rate for particles in a system over a time span very small compared to the actual relaxation time of that system because it is based on measuring statistics of transitions and not actual changes in structure. In practice it suffices to gather transition statistics over several dynamical times, to allow state lifetimes to be properly sampled. This enables one to employ a much larger number of particles in the simulations and numbers realistic for a globular cluster are well within reach of present day supercomputers.

However, it is unclear to what extent the assumption of successive states being uncorrelated is satisfied. McMillan et al. (1988) also studied the properties of the random walk to derive a diffusion coefficient. They used the Fourier transform of the energy time series $E(t_i)$, where single particle energies $E$ were sampled at times $t_i = i\delta t, i = 1, \cdots M$. An advantage of their method is that they can tailor the sampling interval via its Fourier transform to establish that the diffusion limit is reached, i.e., that the long time behaviour $\Delta E^2 \propto t$ characteristic for diffusion is sampled. A disadvantage of their method is that they only measure the second diffusion coefficient $D_2$ and in addition have some uncertainties about which energy the measured $D_2$ should be associated with, since the energy of the particle may change significantly between $t_0$ and $t_M$.

In the following we will consider the identification of states as an *Ansatz* to be born out by further analysis.

The method described here gives wrong results for particles in a bound binary system*. The argument goes as follows: a member of a binary will not have constant energy (unless the eccentricity is zero). Consequently, the method described so far will wrongfully decide that such a particle is undergoing frequent transitions (one per period) but clearly energy changes *are* correlated from one state to the next. Fortunately, dynamically formed binaries should not be a problem in this investigation because they are soft due to the applied softening and hence will not strongly influence measurements of the $D$'s.

### 4.2 The transition probability $\mathcal{P}(E, \Delta E)$

The energy per unit mass $E(t)$ and its first three derivatives as a function of time $t$ are computed during the simulations for all particles using the same high-order method as used to update positions and velocities. Local energy maxima are detected from $dE/dt = 0$, $d^2E/dt^2 < 0$ and are used to define beginning $t_0$ and end $t_1$ of particle states. Elapsed time $\delta t = t_1 - t_0$, initial particle energy $E(t_0)$ and the energy change $\Delta E(\%) \equiv 100 * (E(t_1) - E(t_0))/E(t_0)$ are recorded and stored for each transition. This is done by counting the number of transitions $\sigma(E, \Delta E, \delta t)$ with energy change between $\Delta E$ and $\Delta E + d\Delta E$, from a given state characterised by an initial energy between $E$ and $E + dE$ and a lifetime between $\delta t$ and $\delta t + d\delta t$. Bin boundaries for storing this number of transitions $\sigma$ are chosen as follows: initial energy $E$ is binned linearly from $E_{\min} = -1.5\Psi(0)$ to $E_{\max} = 0.01\Psi(0)$, relative energy change $\Delta E/E$ is binned logarithmically from $0.01\%$ to $50\%$, and state lifetimes are binned logarithmically from $0.1\,P_{\rm hard}$ to $3t_d$. We use 128 bins in $E$, $2 \times 128 + 1$ bins for $\Delta E/E$ (128 for positive energy changes, 128 for negative energy changes and 1 bin for $|\Delta E/E| < 0.01\%$) and 128 bins for $\delta t$. Given $\sigma$, $\mathcal{P}$ can be computed from

$$N(E)\mathcal{P}(E, \Delta E)dEd\Delta E = \frac{1}{T} \sum_{\delta t} \sigma(E, \Delta E, \delta t), \qquad (17)$$

where $dE$ and $d\Delta E$ are bin widths and $T$ is the total simulation time.

---

* I would like to thank S. Aarseth for pointing this out to me.

**Table 1.** Summary of runs

| N/512 | $W_0$ | $T/t_d$ [a] | $\Delta E_T/E_T$ [b] | $\sum \sigma$ [c] |
|---|---|---|---|---|
| 1 | 9 | 4 x (10+4) [d] | 8.2E-5 | 127 855 |
| 4 | 9 | 27+3 [e] | 1.8E-3 | 504 868 |
| 4 | 3 | 27+3 [e] | 1.2E-3 | 297 687 |
| 16 | 9 | 9 | 1.4E-3 | 1 075 069 |
| 64 | 9 | 9 | 2.0E-3 | 7 154 739 |

[a] total simulation time in dynamical times
[b] Relative total energy change over simulation time
[c] Total number of transitions during run
[d] four runs averaged, each system was run over 4 $t_d$ before collecting statistics
[e] integrated over 3 $t_d$ before collecting statistics

### 4.3 Experimental energy diffusion coefficients

Several quantities of interest can be computed as moments of $\sigma$, or alternatively of $\mathcal{P}$, through Eq.(17). The number $N(E)dE$ of particles at the energy $E$ can be computed from:

$$N(E)dE = \frac{1}{T} \sum_{\Delta E, \delta t} \sigma(E, \Delta E, \delta t)\, \delta t, \qquad (18)$$

where $dE$ is the energy bin width. In turn, the diffusion coefficients can be computed from

$$D_n(E)N(E)dE = \frac{1}{T} \sum_{\Delta E, \delta t} \sigma(E, \Delta E, \delta t)\, (\Delta E)^n \qquad (19)$$

Note that these equations are only accurate if the major contribution to the sums on the rhs is due to states with lifetimes $\delta t < \delta t_{\max} = 3 t_d << T$, since only those are sampled properly.

## 5 RESULTS

Table 1 presents a summary of the runs performed. The CPU time required to perform these runs is dominated by the $32k$ model which needed approximately 10 days of CPU time to complete. The time evolution of the Lagrangian radii was investigated to look for obvious signs of evolution in these models. Any systematic changes in these Lagrangian radii are both small with respect to the erratic changes due to low particle number and with respect to the difference in Lagrangian radii between the $W_0 = 3$ and $W_0 = 9$ models.

Figure 1 compares the number of particles $N(E)/N$ obtained from Eq. (18) against the corresponding theoretical curve for the initial King model. The agreement between them is excellent which shows that time averages of the system obtained from the simulations are a good measure for properties of the initial state of the system. We suggest that any small differences are due to initial transients, caused by the fact that the initial models are not completely in equilibrium.

A comparison of the theoretical diffusion rates $N(E)D_n(E)$ ($n = 1, 2$) with the $D$'s from Eqs. (6) and (7) respectively, using the value for the Coulomb logarithm computed from Eq. (11) against the experimental values as defined in Eq. (19) is made in Fig. 2 for the N=32k, $W_0 = 9$ case (but see also section 6.3 on mean field relaxation). The dimensionless $D$'s in this figure are scaled so as to make them independent of $N$. The overall agreement is impressive in view of the fact that there are no free parameters (in particular, the overall normalisation is not free) in either curve or data points, and that the assumptions made in deriving them are very different.

Nonetheless, there are systematic differences between experiment and theory. The experimental $D_2$ is higher for strongly bound particles than its theoretical counterpart by a factor up to two. This can be at least partly understood from the realisation that the theoretical expression is singular at energies close to $-\Psi(0)$ and hence fails to describe the diffusion process there. The reason for the singularity is that in the theoretical derivation there can be no particles with $E < -\Psi(0)$ and hence $D_2(E \to -\Psi(0)) \to 0$. Yet in a *finite* realisation of a King model, as used in the simulations, there can be such tightly bound particles and so there is no reason why $D_2$ should become zero there. A similar argument holds for $D_1$.

The experimental $D$'s are larger than the theoretical ones for particles close to the tidal boundary as well. There can be several reasons for that: (1) the initial model is not sufficiently close to equilibrium due to the introduction of smoothing, and so the effect is purely numerical, (2) the effect is real and is due to collective modes or the influence of simultaneous encounters, not taken into account in the theory, (3) the effect is due to the fact that the effective Coulomb logarithm is larger in the outer parts than the value used in the theoretical expressions, $p_{\max} = r_c$.

Fig. 3 shows the ratios of $D_2$'s for various $N$ to $D_2$ for $N = 32k$, all for $W_0 = 9$, after taking into account the scaling with $N$ suggested by Eq. (7). The shape of the experimental $D_2$ curves is very similar for different $N$'s over most of the energy range and in addition they follow the expected scaling $\propto N/\ln \Lambda$ very well, making these ratios $\approx 1$. Consequently, all $D_2$'s show an increase with respect to the theoretical one for loosely bound particles by a factor up to 1.5, which suggests that this effect already noted for the $N = 32k$ case is real and theory underestimates the diffusion rate in the outer parts of this concentrated ($W_0 = 9$) model. Such an underestimate implies a corresponding underestimate of the evaporation rate for this model.

A comparison between experimental $D_1$'s and their theoretical counterparts for various $N$ is made in Fig. 4 from which it is clear that these experimental values follow very well the theoretical prediction. In particular, there is no obvious sign of $D_1$ being larger in the outer parts, as there was in the $N = 32k$ case. $D_1$ for core particles in the $N = 512$ case fails to track the theoretical curve, unlike the other $N$ models, which might be due to the onset of evolution in this fewer $N$ system.

Finally, Fig. 5 compares $D$'s for a different central potential, $W_0 = 3$ and $N = 2k$. The correspondence for $D_2$ is not as good as in the $W_0 = 9$ case, with the experimental $D_2$ typically 50% too low in the outer parts. The correspondence for $D_1$ is better but the statistics are poorer than in the $W_0 = 9$ case.

The diffusion coefficient $D_2$ is a measure of the energy relaxation time $T_E(E)$ (e.g., Spitzer 1987, Eq. (2-61)),

$$T_E(E) \equiv \langle v^2(E) \rangle_V^2 / D_2(E), \qquad (20)$$

where $\langle v^2(E) \rangle_V = 3q/p$ is the phase-space averaged veloc-ity dispersion of stars with energy $E$. In view of the good correspondence between theoretical and experimental $D_2$'s we find good agreement between theoretical and experimental relaxation rates, although theory slightly underestimates $T_E$ both in the central parts and in the halo for the concentrated model (factor $\approx 1.5 - 2$). The correspondence is slightly worse in the $W_0 = 3$ case.

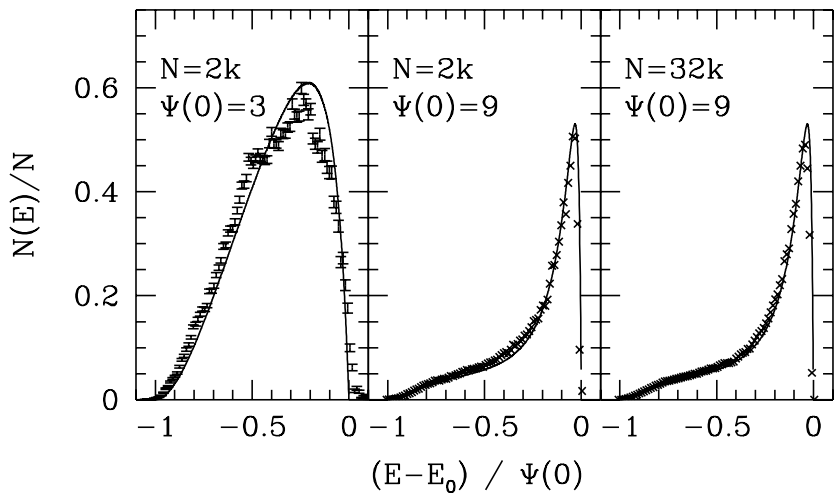

**Figure 1.** Relative number of particles at a given energy $N(E)/N$ obtained from Eq. (18) (symbols) compared against $M(E)/M$, the fractional mass at energy $E$, for a King model (drawn lines) for various $N$ and central concentrations $W_0$, indicated in the figures. Error bars assuming Poissonian errors only are indicated in the $W_0 = 3$ figure. Such errors are smaller than the symbols in the other figures.

## 6 DISCUSSION

In the previous section it was shown that there is in general excellent agreement between the theoretical and experimental diffusion coefficients. In particular, the shape and normalisation of the $D_i(E)$ curves from both approaches agree extremely well (see Fig. 2), using the standard value for the Coulomb logarithm. In addition, the experimental $D$'s follow the theoretically predicted scaling with particle number $N$ very well (see Fig. 3).

However, in spite of this good agreement over most of the energy range, the experimental measurement of $D_2$ is slightly higher (factor $\sim 1.5$) in the outer parts of the concentrated model, although no such deviation is apparent for $D_1$ in those same models (Fig. 4). The measurements of $D_2$ in the outer halo for the same King model with different numbers of particles are consistent amongst themselves, hence, there seems to be a difference between 'experimental' values (based on Eq. (19)) and the 'theoretical' values (from Eqs. (6) & (7)). What causes this small discrepancy? In the following sections we will elaborate on possible reasons. Our conclusion from the discussion below is that (1) since the definition of states does *not* involve any dimensional quantities, it is unlikely to cause a difference between core and halo particles and (2) comparison of the models for different $N$ and different $W_0$ suggests that the (small) deviation is not likely to be due to purely numerical effects like e.g. mean field relaxation. In addition, the theory involves an ad hoc parameter (the Coulomb logarithm) which leads us to suggest that the present form of the theory should not be regarded as the exact value to compare against, or, in other words, the small discrepancy may be an indication that both experimental and theoretical estimates of $D_2$ are approximations to the real diffusion coefficient. Finally, note that the usage of the the Coulomb logarithm does introduce length scales, namely $p_0$ and $p_{\max}$ (section 2.2), so the theory, in contrast to the experiment, does treat core and halo particles on a different footing.

### 6.1 Definition of states

When describing our procedure for measuring the diffusion coefficients (section 4.1) it was mentioned that one should consider the identification of states with the period between two local energy maxima of the single particle as an *Ansatz*. How good is this *Ansatz*? We will first recall some standard results from diffusion theory.

Suppose a particle undergoes a diffusion process (Brownian motion) because it undergoes uncorrelated changes in position $\partial x$ over time scales $\partial t$. Sampling the position $x$ and the associated changes in position $\Delta x$ of this particle at time intervals $\Delta t$, one recovers the standard result that the average distance $L$ the particle will wander from its starting position in a time $T$ grows as $L^2 = T \langle \Delta x^2 / \Delta t \rangle$. This assumes that position changes of the particle are random between successive measurements of its position, i.e., it assumes that $\partial t \ll \Delta t$. However, when the latter inequality is satisfied, the value of the diffusion coefficient $D \equiv \langle \Delta x^2 / \Delta t \rangle$ is *independent* of the sampling interval $\Delta t$.

Next, let us complicate matters and assume that the properties of the position change, $\partial x$, depend on position $x$. To be more specific, assume these properties change significantly when $x$ changes by $\Delta L$. Consequently, the diffusion coefficient itself will depend on $x$, $D = D(x)$. It will be clear that to measure $D(x)$, one has to sample the position of the particle sufficiently often, such that $\sqrt{\Delta x^2} \ll \Delta L$, or, $\Delta t(x) \ll T_d(x) \equiv \Delta L^2/D(x)$. This expresses the fact that the sampling interval $\Delta t(x)$ should be much smaller than the actual characteristic diffusion time $T_d(x)$. Summarising, the sampling interval needs to be such that $\partial t(x) \ll \Delta t(x) \ll T_d(x)$. If this inequality is satisfied, the measured value of $D(x)$ will not dependent on the actual value $\Delta t(x)$ used.

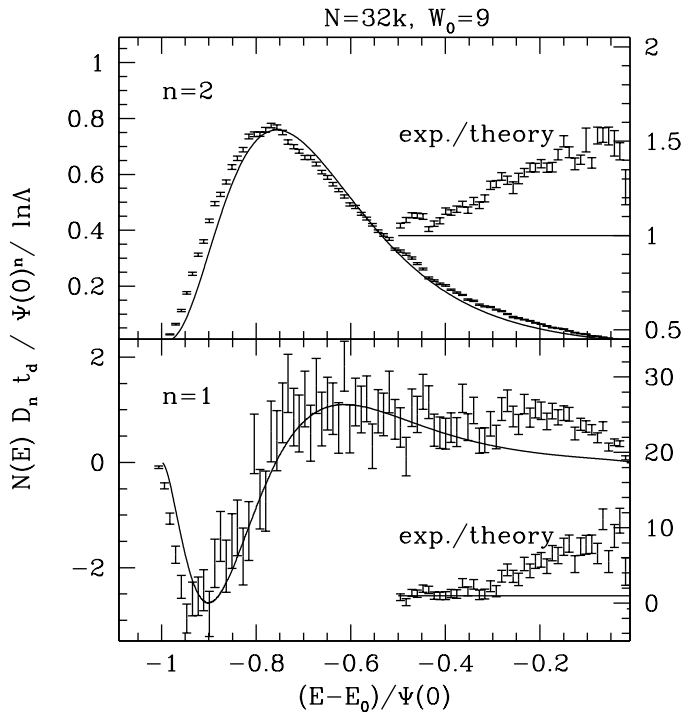 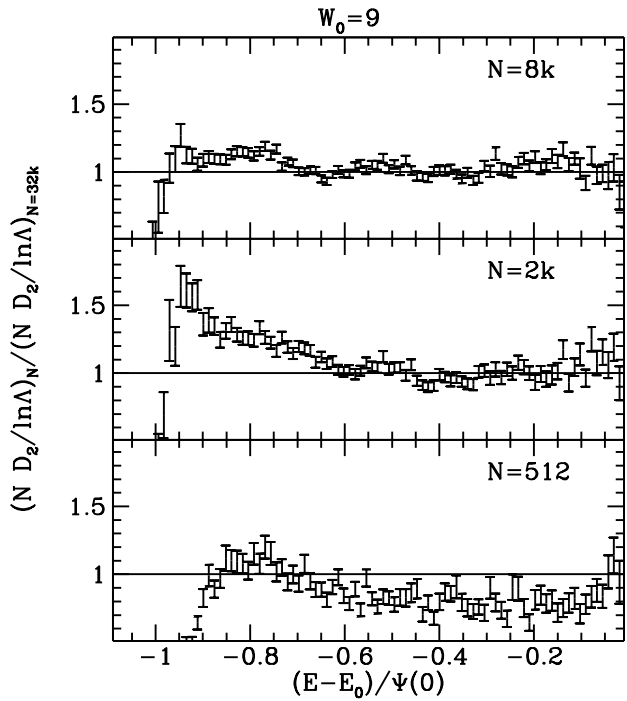

**Figure 2.** Left scales: comparison between theoretical (curves) and experimental (symbols) energy-diffusion rates for the N=32k, $W_0 = 9$ case. The diffusion coefficients are scaled so as to make them independent of $N$. Right scales: idem for the ratio of experimental over theoretical $D$'s for loosely bound particles ($(E - E_0)/\Psi(0) > -0.5$) indicated as 'exp./theory'. Top graph refers to $D_2$, lower graph to $D_1$. Error bars are $1\sigma$ Poissonian errors only.

**Figure 3.** Ratios of diffusion coefficient $ND_2/\ln\Lambda$ for given $N$ over that quantity for $N = 32k$ and $W_0 = 9$, showing the expected scaling $\propto N/\ln\Lambda$. Top to bottom: $N = 8k$, $N = 2k$, $N = 512$.

This example can be applied to energy diffusion in a stellar system. However, there is a caveat: which time interval should be identified with $\partial t$, the duration over which energy changes are correlated? If most of the relaxation is due to near encounters one could presumably take $\partial t \sim \lambda/v$, where $\lambda$ is the (local) interparticle distance and $v$ the (local) velocity dispersion, because the positions of near particles will be uncorrelated when studied intervals of time $\sim \partial t$ apart. However, if relaxation is due to collective processes (e.g. due to interaction of particles with a collective oscillation of the system as occurs during violent relaxation), energy correlations and hence $\partial t$ could be much longer, even comparable to the local dynamical time $\sim 1/\sqrt{\rho}$.

Not much is known about the time correlation of single particle energies in $N$-body systems. In making our definition of particle states as the time between two successive maxima in $E(t)$, we have used the properties of $E(t)$ itself to define when energy changes appear to cease to be correlated. *I have not proven in this paper that this is in fact the case.* This assumption is solely made plausible by suggesting that after $E$ has reached a local maximum, it has 'forgotten' what caused its previous minimum. For example, if a long time scale energy change (an oscillation of the system?) dominates the energy change of the particle, the states definition will naturally select this long time scale for $\partial t$. If, however, most of the energy change is due to the close passage of another particle, the state will be identified as the time over which this other particle has a major influence on the particle under consideration. In addition to this admittedly hand waving justification of the definition, there are some independent points in favour of the choice. Firstly, since there are no dimensional quanti-ties involved, the definition treats all particles (e.g. in core or halo) in *exactly* the same manner. Consequently, even if the inequality $\partial t \ll \Delta t$ were *not* satisfied, it would appear mysterious that the resulting experimental $D_2$'s fit the theoretical ones in the core, but not in the halo. In addition, choosing too small values of $\Delta t$ can only lead to an overestimate of $D_2$, since energy changes are added in quadrature to compute this quantity. Consequently, such an argument does not allow one to explain why the measured values of $D_2$ appear too small compared with the theoretical values for the $W_0 = 3$ model. Secondly, from the measured values of $D_2$ it is clear that the second equality, $\Delta t \ll T_E$ is always satisfied: in fact, if it were not, the process could not be considered a diffusion process in the first place! (Note in passing that it is impossible to satisfy both inequalities with a single (energy independent) time step: indeed, the core relaxation time $T_E(\text{core})$ is actually shorter than the characteristic time scale $\lambda/v$ in the halo, making it impossible for a single $\Delta t$ to satisfy $\partial t(E) \ll \Delta t \ll T_E(E)$ throughout the system.) In addition, note that our definition of states allows us to measure $D_1(E)$ as well. To our knowledge it is the first time that this quantity has been measured directly, and its measured value agrees excellently with the theoretical prediction for all $N$ and $W_0$, both in core and halo.

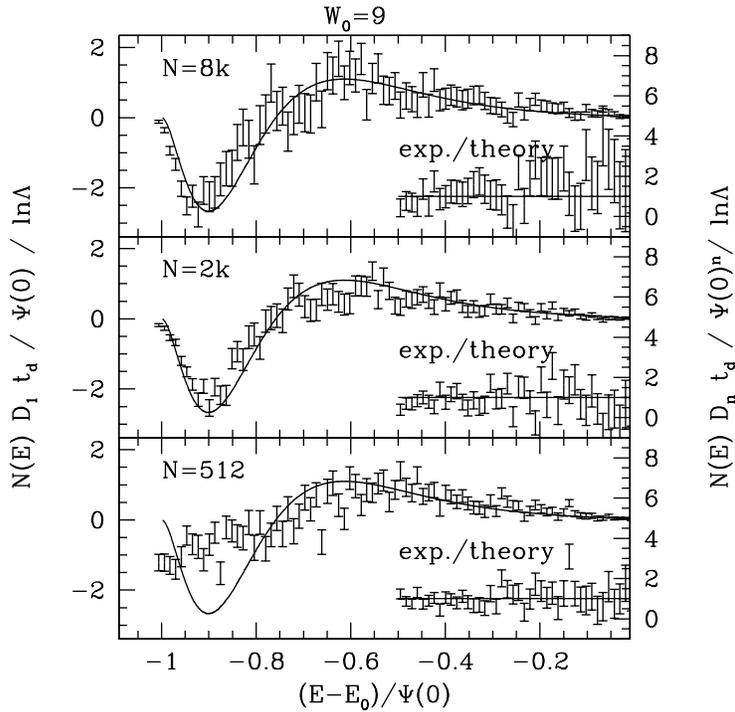

**Figure 4.** Comparison between experimental and theoretical $D_1$'s as in lower panel of Fig. 2 but for different $N$'s indicated in each panel.

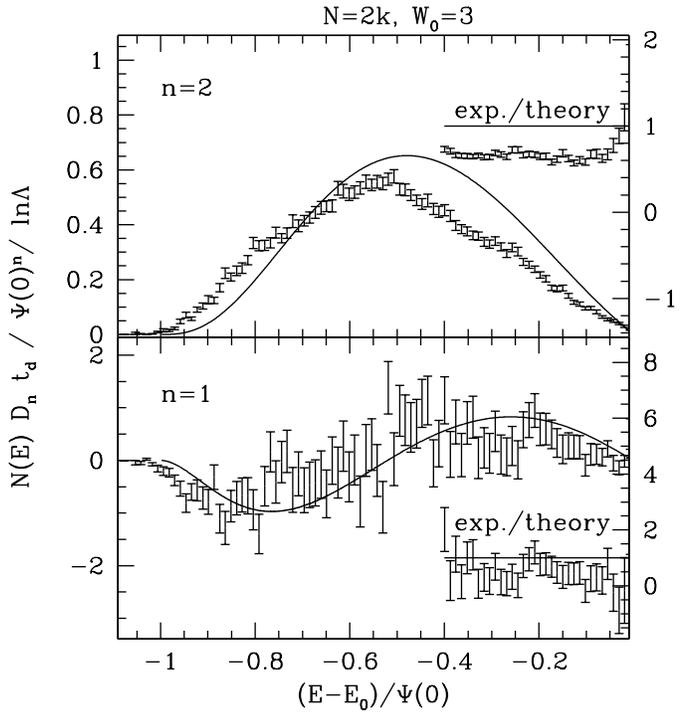

**Figure 5.** Same as Fig.2 but for $W_0 = 3$ and showing ratios for loosely bound particles with $(E - E_0)/\Psi(0) > -0.4$.

Finally, as we remarked before, the theoretical expressions are singular, both in the very core of the system ($E \to \Psi(0)$) and in the outer halo ($E \to E_0$). This is because the theory does not take into account that in an actual $N$-body realisation of the King model, there *can* be particles more strongly bound than $\Psi(0)$ or more loosely bound than $E_0$, unlike in the analytic model. This singular nature of the theoretical $D$'s causes the discrepancy most clearly seen in e.g. Fig. 3 for $E \to \Psi(0)$. Clearly, at least in this case, it is the theoretical expression which is to blame for the discrepancy!

We conclude that, although the assumption that energy changes are uncorrelated between successive states (where states are defined in section 4.1) remains an *Ansatz* until further work is done, the fact that the definition treats halo and core particles in exactly the same way suggests that the small discrepancy between experiment and theory in the halo of the models is unlikely to be due to a failure of the present method.

### 6.2 Simulations

The experimental curves compare the diffusion coefficients of the numerical King model with those of an analytic King model of specified concentration. Is this comparison fair, i.e., is the numerical King model a good representation of the analytic model? In turn, we will discuss the importance of discreteness ($1/\sqrt{N}$ noise), possible out-of-equilibrium effects (breathing modes) and the change in structure as a consequence of the relaxation process itself (leaving the consequences of the *rate* of change of structure due to relaxation effects to the next section).

When generating a King model with a finite number $N$ of particles one naturally suffers from $1/\sqrt{N}$ noise and one would expect the model to be a better representation of the analytic system for larger values of $N$. However, as Fig. 3 testifies, the experimental values of $D_2$ for different $N$ are consistent amongst themselves, over the range 64 in $N$ shown. In addition, such $1/\sqrt{N}$ noise would be unable to explain why experimental $D_2$'s fit in the core (which has *few* particles) and not in the halo.

A numerical model generated using a random number generator is never completely in equilibrium. In the present case, this is aggravated by the fact that the analytical model does not take into account the smoothing employed in the numerical model. Consequently, it might be expected that the numerical King models do not start in equilibrium. However, as in the previous case, the amount by which this influences the results should depend on $N$, since the numerical model actually converges to the analytic one for $N \to \infty$. Consequently, the amount that such breathing modes contribute to the measured value of $D_2$ should be different for different $N$, yet no such effect is seen in Fig. 3: it appears that the non-equilibrium initial state from which these models are started does not overly influence the measurement of $D_2$.

Finally, we have compared numerically evolved models with a specific unevolved King model. But how much has the relaxation process itself changed the numerical models? The half mass relaxation time $T_E(h)$ is known to be of order $N/\ln \Lambda$ times the dynamical time $t_d$ (e.g. Spitzer (1987), BT87), hence, $\approx 10^4 \, t_d$ for the $N = 32k$ model, using the measured value $\ln \Lambda \approx 3$ for the Coulomb logarithm. Consequently, we have evolved that model over only 0.1% of its relaxation time: clearly, relaxation itself is completely unimportant for the $N = 32k$ case and is not able to explain the small discrepancy in the halo where the relaxation time is likely to be even larger than $T_E(h)$. The unimportance of relaxation itself is born out by the behaviour of the Lagrangian radii for this model, which remain virtually identical to the analytic ones over the whole time-span the simulation lasted. In addition, since the relaxation time is a strong function of $N$, the simulated time for the different models, in units of their relaxation time, is very different, e.g., the $N = 512$ model has been integrated over $\sim 20\%$ of its relaxation time (so 200 times longer than the $N = 32k$ model), yet the measured value of $D_2$ agrees well with the measurement from the $N = 32768$ case in the halo.

### 6.3 Mean field relaxation

The rate at which the single particle energy $E(t)$ changes is a sum of the rate due to diffusion, as quantified by the theoretical expression for $D_1$, and due to mean field relaxation, which changes the potential as a function of radius and hence $E(t)$ as well. It is the sum of these two that is measured by the experimental $D_1$. Specifically, $\langle dE/dt \rangle_T = \langle dE/dt \rangle_{MF} + \langle dE/dt \rangle_D$ (T for total, MF for mean field, D for diffusion) and Eq. (2) is a measure of $\langle dE/dt \rangle_D$ but Eq. (19) measures $\langle dE/dt \rangle_T$. An estimate for $\langle dE/dt \rangle_{MF}$ can be obtained using the Fokker-Planck simulations of Cohn (1979, 1980), who gives an estimate of the logarithmic rate of change of the central density, $\rho(0)$, in units of the central relaxation time $tr(0)$:

$$\xi \equiv tr(0) \frac{d \ln(\rho(0))}{dt}. \quad (21)$$

Using Figs. 2 and 3 in Cohn (1979) to convert the central relaxation time to the half-mass relaxation time $tr_h = F \times tr(0)$, with $F \approx 100$ for a $W_0 = 9$ King model, we estimate for the rate of change in the centre:

$$\begin{aligned} |\langle \frac{dE}{dt}(0) \rangle_{MF} \frac{tr_h}{\Psi(0)}| &\approx |\frac{d\Psi(0)}{dt} \frac{tr_h}{\Psi(0)}| = |\frac{d\ln(\Psi(0))}{dt} tr_h| \\ &\approx |\frac{d\ln(\rho(0))}{dt} tr_h| = F\xi \approx 0.5 \end{aligned} \quad (22)$$

using the value $\xi \approx 5 \times 10^{-3}$ (Fig. 6 in Cohn 1980); $tr_h$ is the initial half-mass relaxation time, which corresponds to a Plummer model and so is likely to be larger than for the concentrated King model. (Cohn's simulations start from a Plummer model but he shows that the later stages of evolution can be fit reasonable well by members of the family of King models.) In Fig. 6 we plot $\langle dE/dt \rangle_T = D_1(E)$ obtained from Fig. 2 and converting the dynamical time to half-mass relaxation time using $tr_h = N t_d / 8 \ln \Lambda$ (BT87, Eq. 4-9). Comparing the numerical value in Eq. (22) with Fig. 6 it is clear that the central value $|\langle dE/dt(0) \rangle_{MF}|$ is at least an order of magnitude *smaller* than the central value of $|\langle dE/dt \rangle_T|$. In addition, the large difference between $\langle dE/dt(0) \rangle_{MF}$ and $\langle dE/dt \rangle_T$ suggest that $\langle dE/dt \rangle_T \approx \langle dE/dt \rangle_D$, which explains the good correspondence in Fig. 4. We conclude that the contribution to $\langle dE/dt \rangle_T$ of the mean field relaxation rate is likely to be a small fraction ($\leq 1\%$ say) of the contribution due to diffusion.

We conclude from these remarks that, although the use of states to measure the diffusion coefficients is based on an unproven assumption, there appear to be no serious shortcomings in either method or numerical simulations that would easily explain why the experimentally measured diffusion coefficient should be off in the halo, even though they fit well in the core. We re-iterate however, that we consider the major conclusion to be drawn from Fig. 2 is that there is excellent agreement between experimental and theoretical diffusion coefficients. More work is needed to judge whether the factor $\approx 1.5$ disagreement in the outer halo is due to hidden inaccuracies in the present measurement of $D_2$ or is due to a slight underestimate by the standard theory. In this connection we suggest that the fact that one has to introduce ad hoc a cut-off in an integral appearing in the theoretical derivation (the Coulomb logarithm) suggests that improvements in the theory might be possible. For example, if one introduces a cut-off for distant encounters, it would appear evident that such a cut-off scale should naturally depend on the location of the particle. We also noted in the Introduction that the very concept of a single Coulomb logarithm becomes suspect when dealing with a very inhomogeneous system, as is the case for these concentrated King models.

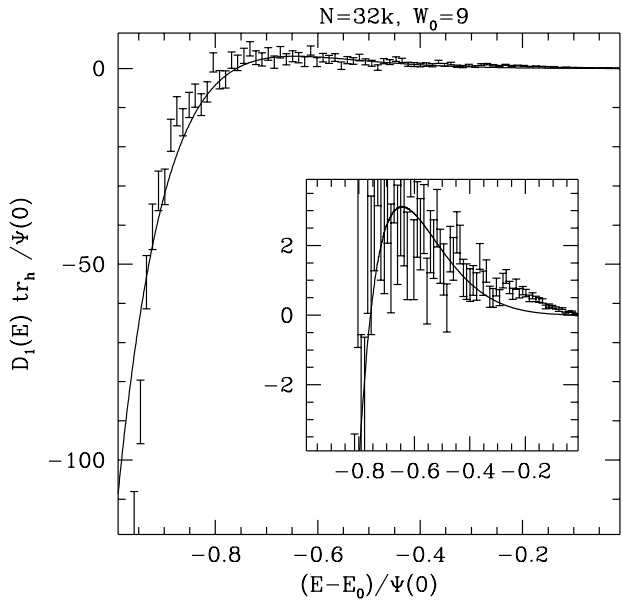

**Figure 6.** Experimental value for $\langle dE/dt\rangle_T \equiv D_1(E)$ from Eq. (18) and (19) (symbols) against the theoretical expression for $\langle dE/dt\rangle_D$ obtained from Eq. (2) (drawn lines) for the concentrated $W_0 = 9$ King model with $N = 32k$ particles. As before, errorbars denote Poissonian errors only. The inset zooms in on the more loosely bound part of the system.

## 7   SUMMARY

Energy diffusion coefficients based on phase-space averages of energy gains per unit time for independent two-body encounters are good approximations to the experimental values obtained from $N$-body simulations. These experimental values are based on the notion of particle states, defined as follows: particle $i$ is considered to be in a given state during the time interval between two consecutive local maxima $E_i(t_b)$ and $E_i(t_e)$ of its energy $v_i^2/2 - \phi_i$. The state is then characterised by its energy $E_i(t_b)$, duration $t_e - t_b$ and transition energy $E_i(t_e) - E_i(t_b)$. It is assumed that different states are independent for all $t_b$ and all particles $i$. Statistics on particle states are then used to compute experimental diffusion coefficients. Although the agreement between these experimental coefficients and their theoretical counterparts is good, theoretical diffusion coefficients are systematically too small by a factor $\approx 1.5 - 2$ both in the core and in the outer halo for the concentrated model, whereas in the less concentrated model theory overestimates the diffusion rate by a similar factor.

Simulations that study the dependence on angular momentum and the effect of a mass-spectrum will be done in the near future. It is important to redo this type of calculation using regularization instead of numerical smoothing to determine whether the results presented here are qualitatively applicable to globular clusters.

## Acknowledgments

George Efstathiou is thanked for discussions and suggestions. Discussions with S. Aarseth, G. Bertin, J. Collett, S. Dutta, D. Heggie, R. Spurzem, M. Stiavelli, E. Vesperini and T.S. van Albada are greatfully acknowledged. D. Chernoff is thanked for tying up some loose ends and J. Binney for extensive comments and suggestions on a previous version of this paper which lead to great improvements. Warm thanks to C. de Loore for advice. All simulations were done at the Connection Machine 2 of the Scuola Normale Superiore di Pisa. This work was begun while I benefited from a fellowship at the SNS di Pisa. I am presently funded by the EEC Human Capital and Mobility Programme under contract CT941463.